\documentclass[12pt]{article}
\catcode`\@=11
\@addtoreset{equation}{section}

\usepackage{amsmath,amssymb,amsfonts,mathrsfs}
\usepackage{graphicx,xcolor,caption,here}
\usepackage{enumerate,comment,cite,url,hyperref}
\usepackage{physics,bm}

\usepackage{needspace}

\global\arraycolsep=2pt
\newcommand{\red}[1]{\textcolor{red}{#1}}

\newcommand{\cF}{\mathcal F}

\newcommand{\cL}{\mathcal L}
\newcommand{\cM}{\mathcal M}

\newcommand{\fJ}{\mathfrak J}
\newcommand{\fL}{\mathfrak L}

\newcommand{\no}{\nonumber}

\newcommand{\pa}{\partial}

\newcommand\al[1]{\begin{align}#1\end{align}}

\newcommand\als[1]{\begin{align}\begin{split}#1\end{split}\end{align}}

\newcommand{\Ad}{ {\rm Ad}}

\allowdisplaybreaks

\begin{document}

\renewcommand{\thefootnote}{\fnsymbol{footnote}}

\begin{flushright} 
RIKEN-iTHEMS-Report-26, STUPP-25-293
\end{flushright}
\vspace*{0.5cm}

\begin{center}
{\Large \bf Courant-Hilbert deformations of \vspace*{0.2cm}\\ Yang-Baxter sigma models
}
\vspace*{1.5cm} \\
{\large  Osamu Fukushima\footnote{E-mail:~osamu.fukushima$\_$at$\_$riken.jp}, 
Takaki Matsumoto\footnote{E-mail:~takaki-matsumoto$\_$at$\_$ejs.seikei.ac.jp}
and Kentaroh Yoshida\footnote{E-mail:~kenyoshida$\_$at$\_$mail.saitama-u.ac.jp
}} 
\end{center}

\vspace*{0.4cm}

\begin{center}
$^{\ast}${\it iTHEMS, RIKEN, Wako, Saitama 351-0198, Japan}
\end{center}

\begin{center}
$^{\dagger}${\it Seikei University, 
3-3-1 Kichijoji-Kitamachi, Musashino-shi, Tokyo 180-8633, Japan}
\end{center}

\begin{center}
$^{\ddagger}${\it Graduate School of Science and Engineering, Saitama University, 255 Shimo-Okubo, Sakura-ku, Saitama 338-8570, Japan}
\end{center}

\vspace{1cm}

\begin{abstract}
We present integrable deformations of Yang-Baxter (YB) sigma models based on the Courant-Hilbert (CH) construction. To this end, we employ the four-dimensional Chern-Simons theory, in which the CH construction is shown in arXiv:2509.22080. As a result, the CH construction works in an intricate way alongside the YB deformations. Remarkably, the resulting deformed action can also be expressed as the sum of the master formula Lagrangian and the trace of the energy-momentum tensor. This result indicates the universality of the correction term. 

\end{abstract}

\setcounter{footnote}{0}
\setcounter{page}{0}
\thispagestyle{empty}

\newpage

\tableofcontents

\renewcommand\thefootnote{\arabic{footnote}}

\section{Introduction}

In the study of integrable systems, one of the important issues is to find a systematic way to describe integrable deformations. One example is Yang-Baxter (YB) deformations of non-linear sigma models in two dimensions    \cite{Klimcik:2002zj,Klimcik:2008eq,Delduc:2013fga,Matsumoto:2015jja} (For details, see a little book \cite{Yoshida}). Each of the YB deformations is characterized by a (skew-symmetric) classical $r$-matrix satisfying the (modified) classical YB equation ((m)CYBE). This is the origin of the name YB deformation, and the YB-deformed sigma model is abbreviated as the YB sigma model. The YB sigma models are classically integrable in the sense of the existence of the Lax pair. The YB deformations are generalized to the AdS$_5\times$S$^5$ superstring case \cite{Delduc:2013qra,Kawaguchi:2014qwa}.  

\medskip 

Recently, another systematic way to describe integrable deformations based on the Courant-Hilbert (CH) construction was presented in \cite{Babaei-Aghbolagh:2025uoz,Babaei-Aghbolagh:2025hlm}. Each of the deformations is characterized by a function that satisfies a partial differential equation (PDE) to maintain the flatness condition of the Lax pair. The general solution to the PDE is given by Courant and Hilbert \cite{Courant} and is characterized by a scalar function that satisfies a boundary condition. In other words, given a scalar function that satisfies a boundary condition, an integrable deformation is specified. We refer to the integrable deformations based on the CH construction as the CH deformations. The CH deformations are associated with operators composed of the energy-momentum tensors. For example, the CH deformations include the $T\bar{T}$-deformation \cite{Smirnov:2016lqw,Cavaglia:2016oda} and the root $T\bar{T}$-deformation \cite{Rodriguez:2021tcz,Babaei-Aghbolagh:2022uij,Ferko:2022cix,Babaei-Aghbolagh:2022leo,Tempo:2022ndz}. 

\medskip 

It is a nice issue to consider a unification of the YB deformation and the CH deformation. A direct approach, while possible in principle, seems very difficult and quite complicated. Hence, to this end, we employ the four-dimensional Chern-Simons (4D CS) theory, which is considered a unified theory of integrable systems \cite{Costello:2019tri}. The action contains a meromorphic one-form. By choosing it and imposing a boundary condition for the gauge field, an integrable sigma model is derived. The prescription to describe the YB deformations within the context of the 4D CS theory was initially presented in \cite{Delduc:2019whp} and was generalized to the homogeneous YB deformations of the AdS$_5\times$S$^5$ superstrings \cite{Kawaguchi:2014qwa} in the work  \cite{Fukushima:2020dcp}. For other works related to the 4D CS theory, see~\cite{Schmidtt:2019otc,Fukushima:2020kta,Tian:2020ryu,Tian:2020pub,Fukushima:2020tqv,Lacroix:2020flf,Caudrelier:2020xtn,Fukushima:2021eni,Stedman:2021wrw,Fukushima:2021ako,Liniado:2023uoo,Berkovits:2024reg}. 

\medskip

In our previous work \cite{Fukushima:2025tlj}, we considered the CH deformations of the 2D principal chiral model (PCM) within the framework of the 4D CS theory. The CH deformations can be incorporated by changing the pole structure of the Lax ansatz. Remarkably, it was shown that when the deformation parameter is dimensionful, the master formula should be corrected by the trace of the energy-momentum tensor. This is consistent with the result \cite{Sakamoto:2025hwi} for the $T\bar{T}$-deformation based on another approach. 

\medskip 

As another aspect of the work \cite{Fukushima:2025tlj}, it is useful to comment on the relation to the auxiliary field sigma model (AFSM) \cite{Ferko:2023wyi,Ferko:2024ali}. The action of the AFSM contains auxiliary fields and an arbitrary potential function. By specifying a form of the potential function, an integrable deformation is determined. Notably, the $T\bar{T}$-deformation and the root 
$T\bar{T}$-deformation are included. The constraints obtained from the equations of motion for the auxiliary fields can be derived from the flatness condition in the CH construction. Then, one of them leads to the PDE that can be solved using the CH method. In this sense, the AFSM is closely related to the CH construction. A related advance is the work of \cite{Fukushima:2024nxm}, which extended the 4D CS theory by including auxiliary fields and an arbitrary potential function in order to derive the AFSM. The AFSM itself has been studied in various directions, including applications to T-duality \cite{Bielli:2024khq}, integrable higher-spin deformations \cite{Bielli:2024ach,Bielli:2025uiv}, (bi)-YB deformations \cite{Bielli:2024fnp}, and deformations of (semi-)symmetric space sigma models \cite{Bielli:2024oif,Ferko:2025bhv,Cesaro:2024ipq}. These developments may be helpful in generalizing the CH construction.

\medskip 

In this paper, we will incorporate the YB deformation into the study of \cite{Fukushima:2025tlj}. Although the generalized Lax ansatz is the same as in \cite{Fukushima:2025tlj}, the resulting Lax form contains a deformed current associated with the YB deformation. As a result, the CH deformation, together with the YB deformation, is realized in a nontrivial way. It should be remarked that the master formula Lagrangian needs to be corrected by the trace of the energy-momentum tensor. This is the same as in \cite{Fukushima:2025tlj} and indicates the universality of the correction term. 

\medskip 

This paper is organized as follows. Section 2 is a brief introduction to the 4D CS theory. In section 3, we present the generalized Lax ansatz for the homogeneous YB deformation. As a result, the master formula is evaluated to a simple form. In section 4, we carry out the same analysis as in section 3 for the YB deformation based on the mCYBE. Remarkably, the same expression of the master formula is obtained. In section 5, we derive the true Lagrangian, which leads to the correct equations of motion. It contains the trace of the energy-momentum tensor, as in the case with no YB deformation. In section 6, we discuss how to perform the CH construction. In particular, it is necessary to remove the deformed current from the solutions obtained by the CH method. Indeed, we do this for two concrete examples: the root $T\bar{T}$-deformation and the $T\bar{T}$-deformation. Section 7 is devoted to the conclusion and discussion. In Appendix A, the energy-momentum tensor for the true Lagrangian is computed. Appendix B explains the derivation of (\ref{++--scalars}) in detail.   

\Needspace{8\baselineskip}
\paragraph{NOTE added:} In the course of preparing this paper, we became aware of an important study \cite{Baglioni:2025tsc}, in which the equivalence between the CH construction and the auxiliary field method has been demonstrated. Furthermore, the YB deformation is also studied, which overlaps with our paper in some respects. 
In \cite{Baglioni:2025tsc}, the equivalence is shown by introducing the $\mu$-frame formulation of the AFSM. The $\mu$-frame YB-deformed AFSM is also obtained, but it is nontrivial whether the model admits a CH-like formulation after integrating out the auxiliary field. By contrast, our approach directly leads to the CH-deformation of the YB sigma model.

\section{A brief introduction to the 4D CS theory}

To study the CH deformations of the YB sigma model, we shall employ the 4D CS theory, from which 2D integrable sigma models can be obtained systematically \cite{Costello:2019tri}. 

\medskip 

The 4D CS theory is a four-dimensional gauge theory with a Lie group $G$\,. The gauge field $A$ lives on a 4D manifold $\cM\times \mathbb{C}P^1$\,, where $\cM$ denotes 2D Minkowski spacetime with coordinates $x^\mu=(\tau,\sigma)$ and metric $\eta_{\mu\nu}=\operatorname{diag}(-1,1)$. We also use the light-cone coordinates defined by $\sigma^\pm := (\tau\pm\sigma)/2$\,.
The complex projective space $\mathbb{C}P^1$ is equipped with a complex coordinate $z$\,. 

\medskip 

The classical action for $A$ is given by
\begin{align}
    S[A] & = \frac{i}{4\pi}\int_{\cM\times\mathbb{C}P^1}\omega\wedge CS(A)\,, 
    \label{4dCS-action}
    \\  
   CS[A] & := \tr(A\wedge dA+\frac{2}{3}A\wedge A \wedge A) \,, 
\end{align}
where $\omega$ is a meromorphic one-form defined as 
\begin{align}
    \omega := \varphi(z)\,dz
\end{align}
with a meromorphic function $\varphi(z)$ on $\mathbb{C}P^1$\,. Since $\omega$ is complex, to ensure the reality of the action, the Lie group $G$ should be complexified as $G^{\mathbb{C}}$ and then the gauge field $A$ takes a value in the complexified Lie algebra $\mathfrak{g}^{\mathbb{C}}$ as $A:\cM\times\mathbb{C}P^1 \to\mathfrak{g}^{\mathbb{C}}$\,. 
For the details, see \cite{Delduc:2019whp}. 
Note that the $z$-component of $A$ is taken to be zero, i.e., $A_z=0$ by using the extra gauge symmetry under $A\mapsto A + \chi\, dz $\,.

\medskip

By varying the action with respect to $A$ as $A\mapsto A+\delta A$\,, we can obtain the following equations of motion (EOMs)
\begin{align}
    0=&\,\omega\wedge \tr(A\wedge dA + A\wedge A)\,,
    \label{bulk-A}
    \\
    0=&\,d\omega \wedge \tr(\delta A\wedge A)\,.
    \label{boundary-A}
\end{align}
We refer to the first equation \eqref{bulk-A} as the bulk EOM. The second one \eqref{boundary-A} is referred to as the boundary EOM since it has support only at the poles of $\varphi(z)$\,.

\medskip

Let us now introduce a Lax form $\fL$ through a formal gauge transformation
\begin{align}
    A = -d\hat{g}\,\hat{g}^{-1} +\hat{g}\,\fL\, \hat{g}^{-1}\,
    \qquad
    \mbox{s.t.}~~
    \fL_{\bar{z}}
    =0\,,
\end{align}
with a smooth group element $\hat{g}:\cM\times\mathbb{C}P^1\to G^{\mathbb{C}}$\,. In terms of the Lax form, the bulk EOM \eqref{bulk-A} is rewritten as 
\begin{align}
    0=&\,\pa_{+}\fL_--\pa_{-}\fL_+ + [\fL_+,\fL_-]\,,
    \label{Lax-flat}
    \\
    0=&\,\varphi(z)\,\pa_{\bar{z}}\fL_+ = \varphi(z)\,\pa_{\bar{z}}\fL_-\,.
    \label{Lax-meromorphic}
\end{align}
The equation \eqref{Lax-meromorphic} indicates that $\fL_{\pm}$ may have non-trivial pole structure at the zeros of $\varphi(z)$\,, and the pole structure of $\fL$ should match the zero structure of $\varphi(z)$\,. 

\medskip 

By performing the complex integral over $\mathbb{C}P^1$ \cite{Benini:2020skc}, the action can be rewritten as 
\begin{align}
S[\{g_x\}_{x\in \mathfrak{p}}] = \frac{1}{2}\sum_{x\in \mathfrak{p}}\int_{\mathcal{M}} {\rm tr}\left( {\rm res}_x(\varphi\, \mathfrak{L})\wedge g_x^{-1}dg_x
\right) + \frac{1}{2}\sum_{x\in \mathfrak{p}} ({\rm res}_{x}\varphi)\int_{\mathcal{M}\times I} I_{\rm WZ}[g_x]\,,
\label{master-formula}
\end{align}
where $\mathfrak{p}$ is the set of poles of $\varphi(z)$ and $g_x := \hat{g}|_{z=x}$\,. This is called the master formula~\cite{Delduc:2019whp}. The second term in \eqref{master-formula} is not relevant in the following. For the details of $I_{\text{WZ}}$\,, see~\cite{Delduc:2019whp}.

\section{Generalizing Homogeneous YB deformations}

The homogeneous YB deformations can be described by changing boundary conditions of the gauge field $A$ \cite{Delduc:2019whp}. In the following, we will focus on the 2D PCM case and generalize the Lax form to accommodate the CH construction by following the work \cite{Fukushima:2025tlj}.

\subsection{Homogeneous YB deformations}

We begin with a meromorphic function for 2D PCM, 
\begin{align}
    \varphi(z) =&\, \frac{1-z^2}{z^2}\,. \label{PCM}
\end{align}
Then, in order to include the homogeneous YB deformations, let us impose the following boundary condition \cite{Delduc:2019whp} at the poles of $\varphi(z)$ in (\ref{PCM})\,,  
\begin{align}\begin{split}
    A|_{z=\infty}=&\,0\,,
    \\
    A|_{z=0} =&\, \eta \, R\big(\pa_{z}A|_{z=0}\big)\,, 
\end{split}\label{bc-YB}
\end{align}
which solves the boundary EOM \eqref{boundary-A}. Here $\eta$ is a real constant parameter, and $R$ is a skew-symmetric $R$-operator satisfying the homogeneous CYBE 
\begin{align}
    \big[R(\mathsf{x}),R(\mathsf{y})\big] - 
    R\big([R(\mathsf{x}),\mathsf{y}] + [\mathsf{x},R(\mathsf{y})]\big)
    =0\,,
    \qquad
    {}^\forall\mathsf{x}\,,\mathsf{y}\in \mathfrak{g}^{\mathbb{C}}\,. \label{hYB}
\end{align}

\subsection{Generalized Lax form}

As for the ansatz of the Lax form, which is consistent with the zero structure of $\varphi(z)$\,, let us consider the following one \cite{Fukushima:2025tlj}, 
\begin{align}
    \fL_{\pm}= \frac{V_{\pm}+zK_{\pm}}{1-z^2} +U_{\pm}\,,
    \label{ansatz-WZ}
\end{align}
where $V_{\pm}\,,$ $K_{\pm}$ and $U_{\pm}$ are $\mathfrak{g}^{\mathbb{C}}$-valued smooth functions on $\cM$\,. In comparison to the ansatz proposed in \cite{Delduc:2019whp}, 
the additional functions $K_{\pm}$ are included, and especially the pole structure is different. When $K_{\pm} = \pm V_{\pm}$\,, the ansatz in \cite{Delduc:2019whp} is reproduced. In this sense, the inclusion of $K_{\pm}$ can be regarded as performing deformations of the seed theory specified by $\varphi(z)$ in \eqref{PCM}. 

\medskip 

With the residual gauge symmetry, we may take $\hat{g}$ at the poles of $\varphi(z)$ as 
\begin{align}
    \hat{g}|_{z=0}= g_0 =:g\,,\qquad \hat{g}|_{z=\infty}=\bm{1}\,. 
\end{align}
Then, by using the boundary condition (\ref{bc-YB})\,, the Lax ansatz \eqref{ansatz-WZ} is evaluated as 
\begin{align}
    \fL_{\pm}=
    \frac{j_{\pm}+\eta\, R_{g}K_\pm +z K_{\pm}}{1-z^2}\,, \qquad j_{\pm} := g^{-1}\partial_{\pm}g\,.
    \label{ansatz-WZ-2}
\end{align}
Here $R_g$ is a chain of the operations defined as $R_g :=\Ad_{g^{-1}}\circ R \circ \Ad_g$\,, where Ad$_g$ is the adjoint operation with a group element $g$\,, defined as ${\rm Ad}_g(X) := g X g^{-1}$ for any $X \in \mathfrak{g}^{\mathbb{C}}$\,.

\subsection{The flatness of Lax form}

For later convenience, it is helpful to introduce the following quantities
\begin{align}
    \fJ_{\pm} := &\, \pm K_{\pm}\,,
\label{current}    \\
    \tilde{j}_{\pm} := &\, j_{\pm} + \eta R_g K_{\pm}
    = j_{\pm} \pm \eta R_g \fJ_{\pm}\,.
\label{tilde-def}
\end{align}
Note that the undetermined functions are now represented by $\mathfrak{J}_{\pm}$\,. Since the current $\tilde{j}$ depends on the $R$-operator, we refer to it as the deformed current. 

\medskip 

Then, the Lax ansatz \eqref{ansatz-WZ-2} is rewritten as 
\begin{align}
    \fL_{\pm}=&\,
    \frac{\tilde{j}_{\pm} \pm z\, \fJ_{\pm}}{1-z^2}\,.
    \label{YB-Lax}
\end{align}
This expression is apparently the same as the one for the YB-deformed AFSM \cite{Bielli:2024fnp}, in which the current components $\fJ_{\pm}$ include the auxiliary fields.

\medskip 

Let us examine the flatness condition of $\fL_{\pm}$ in (\ref{YB-Lax})\,. Since $\fL_{\pm}$ contain undetermined functions $K_{\pm}$\,, it is not trivial whether $\fL_{\pm}$ satisfy the flatness condition or not. 

\medskip 

By using the expression (\ref{YB-Lax})\,, we obtain that 
\begin{align}
&     \pa_{+}\fL_- - \pa_{-}\fL_+ + [\fL_+,\fL_-] \notag \\ 
    =&\,
	\frac{\pa_+\tilde{j}_- - \pa_-\tilde{j}_++[\tilde{j}_+,\tilde{j}_-] -z(\partial_+\mathfrak{J}_- + \partial_-\mathfrak{J}_+)}{1-z^2}
    \no\\
    &
	+ \frac{ z([\mathfrak{J}_+,\tilde{j}_-] - [\tilde{j}_+,\mathfrak{J}_-]) -
	z^2([\mathfrak{J}_+,\mathfrak{J}_-] - [\tilde{j}_+,\tilde{j}_-])}{(1-z^2)^2}\,. 
\end{align}
Hence, in order to satisfy the flatness condition, 
the following conditions should be satisfied: 
\begin{align}
    \pa_{+}\fJ_- + \pa_{-}\fJ_+ =& 0\,,
    \label{J-conservation}
    \\
    \pa_{+}\tilde{j}_- - \pa_{-}\tilde{j}_+ + [\tilde{j}_+,\tilde{j}_-]=& 0\,,
    \label{flatness-tilde} \\ 
    [\tilde{j}_{+},\fJ_{-}] =&\, [\fJ_{+},\tilde{j}_-]\,,
    \label{tilde-comm1}
    \\
    [\tilde{j}_+,\tilde{j}_-]=&\,[\fJ_+,\fJ_-]\,.
    \label{tilde-comm2}
\end{align}
Based on the standard knowledge about the Lax pair, the first equation should be the on-shell condition, and it is nothing but the equation of motion. The second equation is the flatness condition for $\tilde{j}$\,, and it can be rewritten as
\begin{align}
    0=&\,\pa_{+}\tilde{j}_- - \pa_{-}\tilde{j}_+ + [\tilde{j}_+,\tilde{j}_-]
    \no\\
    =&\,
    \pa_{+}j_- - \pa_{-}j_+ + [j_+,j_-]
    -\pa_{+}(\eta R_g \fJ_-) - \pa_{-}(\eta R_g \fJ_+) 
    -\big[ \eta R_g\fJ_+, \eta R_g \fJ_- \big] \notag \\ 
 &   -[j_+, \eta R_g\fJ_-]
    +[\eta R_g \fJ_+, j_-]
    \no\\
    =&\,
    -\eta R_g\big( \pa_{+}\fJ_- + \pa_{-}\fJ_+ \big)
    -\eta R_g \big( [j_+,\fJ_-]+[j_-,\fJ_+]\big)
    -\big[\eta R_g \fJ_+,\eta R_g \fJ_- \big]
    \no\\
    =&\,
    -\eta R_g\big( \pa_{+}\fJ_- + \pa_{-}\fJ_+ \big)
    -\eta R_g \big( [j_+,\fJ_-]+[j_-,\fJ_+]\big)
    -\eta R_g\big(\big[\fJ_+,\eta R_g \fJ_- \big] 
    + [\eta R_g\fJ_+, \fJ_- \big]\big)
    \no\\
    =&\,
    -\eta R_g\big( \pa_{+}\fJ_- + \pa_{-}\fJ_+ \big)
    -\eta R_g \big( [\tilde{j}_+,\fJ_-]+[\tilde{j}_-,\fJ_+]\big)\,,
    \end{align}
where we have utilized the following relation
    \begin{align}
        \pa_{\pm}\big(R_g \fJ_{\mu}\big) =&\,
        [R_g\fJ_{\mu},j_{\pm}] + R_g\big([j_{\pm},\fJ_{\mu}]\big) + R_g\pa_{\pm}\fJ_{\mu}
    \end{align}
and the off-shell flatness condition of $j$ on the second line, and the CYBE \eqref{hYB} on the third line. After all, the flatness condition for $\tilde{j}$ in \eqref{flatness-tilde} is satisfied when the commutation relation \eqref{tilde-comm1} and the equation of motion \eqref{J-conservation} are satisfied. In this sense, the flatness condition \eqref{flatness-tilde} is also the on-shell condition. 

\medskip 

The third and fourth conditions (\ref{tilde-comm1}) and (\ref{tilde-comm2}) should be the off-shell conditions as in the case with no YB deformation \cite{Fukushima:2025tlj}. 

\subsection{The master formula}

Finally, the master formula \eqref{master-formula} is evaluated as 
\begin{align}
    S_\eta[g] =-\int_{\cM}\!\! d\sigma^+\wedge d\sigma^-\tr (j^\mu\,\fJ_\mu ) =\frac{1}{2}\int_{\cM}\!\! d\sigma^+\wedge d\sigma^- \tr(j_-\,\fJ_+ + j_+\,\fJ_-)\,. 
    \label{YB-action}
\end{align}
This is apparently the same form as in \cite{Fukushima:2025tlj}. The contribution of the YB deformation appears only through $\mathfrak{J}_{\mu}$\,. 


\section{Generalizing YB deformations}

In the previous section, we have considered the homogeneous YB deformations by changing the boundary conditions of the gauge field $A$\,. Here, we will consider the YB deformation based on the mCYBE
\begin{align}
    \big[R(\mathsf{x}),R(\mathsf{y})\big] - 
    R\big([R(\mathsf{x}),\mathsf{y}] + [\mathsf{x},R(\mathsf{y})]\big)
    +c^2[\mathsf{x},\mathsf{y}]=0\,,
    \qquad
    {}^\forall\mathsf{x}\,,\mathsf{y}\in \mathfrak{g}^{\mathbb{C}}\,,
    \label{mod-YB}
\end{align}
where the $c=i$ and $c=1$ cases correspond to the split and non-split types, respectively. The $c=0$ case corresponds to the homogeneous case that has already been addressed. 

\subsection{The YB deformations}

To consider the YB deformation based on the mCYBE, let us start with the twist function 
\begin{align}
    \varphi(z) =&\,
    \frac{1}{1-c^2\eta^2}\frac{1-z^2}{z^2-c^2\eta^2}\,,
\end{align}
where $c \neq 0$\,, and $\eta$ is a real deformation parameter. 

\medskip 

By following \cite{Delduc:2019whp}, the boundary conditions for the gauge field $A$ at the poles of $\varphi$\,, $z=\infty$ and $z = \pm c \eta$  are taken as follows: 
\begin{align}
    (A|_{z=+c\eta},A|_{z=-c\eta})\in
    \{((R+c)\mathsf{x},(R-c)\mathsf{x})\,|\,\mathsf{x}\in\mathfrak{g}\}\,.
    \label{bc-algebra}
\end{align}
Here, $R$ is a skew-symmetric $R$-operator that satisfies the mCYBE (\ref{mod-YB})\,. 
The condition \eqref{bc-algebra} can be rewritten as 
\begin{align}
    (R-c)A|_{z= +c\eta} = (R+c)A|_{z= -c\eta}\,.
\end{align}
By utilizing the gauge redundancy and the boundary algebra, we can fix $\hat{g}$ as 
\begin{align}
    \hat{g}|_{z=\pm c\eta} = g_{\pm c\eta} =: g\,,\qquad
    \hat{g}|_{z=\infty}=\bm{1}\,.
\end{align}

\subsection{Generalized Lax form}

The same Lax ansatz (\ref{ansatz-WZ}) leads to 
\begin{align}
    V_{\pm}=(1-c^2\eta^2)\,j_{\pm} + \eta\, R_g\, K_{\pm}\,,
    \qquad
    U_{\pm}=0\,.
\end{align}
Then, it is evaluated as 
\begin{align}
    \fL_{\pm}=&\,
    \frac{(1-c^2\eta^2)\,j_{\pm}+ \eta \,R_g\, K_{\pm} +z\, K_{\pm}}{1-z^2}\,.
    \label{ansatz-WZ-2-with-c}
\end{align}

\medskip 

Let us define the currents $\fJ_{\pm}$ and $\tilde{j}_\pm$ as, respectively, 
\begin{align}
     \fJ_{\pm} :=&\, \pm \frac{K_{\pm}}{1-c^2\eta^2}\,,
 \label{current2}   \\
    \tilde{j}_{\pm}:=&\, \frac{(1-c^2\eta^2)\,j_{\pm} + \eta R_g\, K_{\pm}}{1-c^2\eta^2}
    =j_{\pm} \pm \eta R_g \,\fJ_{\pm}\,. 
\label{tilde-def-eta}
\end{align}
Note that the undetermined functions $K_{\pm}$ are represented by $\mathfrak{J}_{\pm}$ again. Then the Lax ansatz \eqref{ansatz-WZ-2-with-c} can be rewritten as
\begin{align}
    \fL_{\pm}=&\, \big(1-c^2\eta^2\big)\,
    \frac{\tilde{j}_{\pm}\pm z\,\fJ_{\pm}}{1-z^2}\,.
    \label{eta-Lax}
\end{align}
The next task is to examine the flatness condition. 

\subsection{The flatness of Lax form}

The flatness of the Lax form \eqref{eta-Lax} holds if the following conditions are satisfied: 
\begin{align}
\pa_{+}\fJ_- + \pa_{-}\fJ_+\,=&\, 0\,,
    \label{J-conservation-eta}
    \\
  \pa_{+}\tilde{j}_- - \pa_{-}\tilde{j}_+ + (1-c^2\eta^2)[\tilde{j}_+,\tilde{j}_-]\,  =&\, 0\,, 
    \label{flatness-tilde-eta} \\ 
    [\tilde{j}_{+},\fJ_{-}] =&\, [\fJ_{+},\tilde{j}_-]\,,
    \label{tilde-comm1-eta}
    \\
    [\tilde{j}_+,\tilde{j}_-]=&\,[\fJ_+,\fJ_-]\,. 
    \label{tilde-comm2-eta} 
\end{align}
The first equation indicates the equation of motion based on standard knowledge about the Lax pair. The second equation is the flatness condition\footnote{The ugly factor $1-c^2\eta^2$ can be removed by rescaling $\tilde{j}_{\pm}$ and so this is the usual flatness condition.} of $\tilde{j}_{\pm}$\,, and it can be rewritten as
\begin{align}
    0=&\,\pa_{+}\tilde{j}_- - \pa_{-}\tilde{j}_+ + (1-c^2\eta^2)[\tilde{j}_+,\tilde{j}_-]
    \no\\
    =&\,
    \pa_{+}j_- - \pa_{-}j_+ + (1-c^2\eta^2)[j_+,j_-]
    -\pa_{+}(\eta R_g \fJ_-) - \pa_{-}(\eta R_g \fJ_+) 
    \no\\
    &-(1-c^2\eta^2)\big[ \eta R_g\fJ_+, \eta R_g \fJ_- \big]
    -(1-c^2\eta^2)[j_+, \eta R_g\fJ_-]
    +(1-c^2\eta^2)[\eta R_g \fJ_+, j_-]
    \no\\
    =&\,
    -\eta R_g\big( \pa_{+}\fJ_- + \pa_{-}\fJ_+ \big)
    -\eta R_g \big( [j_+,\fJ_-]+[j_-,\fJ_+]\big)
    \no\\
    &-c^2\eta^2[j_+,j_-]
    -(1-c^2\eta^2)\big[\eta R_g \fJ_+,\eta R_g \fJ_- \big]
    +c^2\eta^2[j_+, \eta R_g\fJ_-]
    -c^2\eta^2[\eta R_g \fJ_+, j_-]
    \no\\
    =&\,
    -\eta R_g\big( \pa_{+}\fJ_- + \pa_{-}\fJ_+ \big)
    -\eta R_g \big( [j_+,\fJ_-]+[j_-,\fJ_+]\big)
    -\eta R_g\big(\big[\fJ_+,\eta R_g \fJ_- \big] 
    + [\eta R_g\fJ_+, \fJ_- \big]\big)  \no\\ 
    & +c^2\eta^2[\fJ_+,\fJ_-] -c^2\eta^2[j_+,j_-]
    +c^2\eta^2\big[\eta R_g \fJ_+,\eta R_g \fJ_- \big] \no \\ 
    & 
    +c^2\eta^2[j_+, \eta R_g\fJ_-]
    -c^2\eta^2[\eta R_g \fJ_+, j_-]
    \no\\
    %
    =&\,
    -\eta R_g\big( \pa_{+}\fJ_- + \pa_{-}\fJ_+ \big)
    -\eta R_g \big( [\tilde{j}_+,\fJ_-]+[\tilde{j}_-,\fJ_+]\big)
    +c^2\eta^2\big([\fJ_+,\fJ_-]- [\tilde{j}_+,\tilde{j}_-]
    \big)
    \,,
    \end{align}
Here we have utilized the relation,
    \begin{align}
        \pa_{\pm}\big(R_g \fJ_{\mu}\big) =&\,
        [R_g\fJ_{\mu},j_{\pm}] + R_g\big([j_{\pm},\fJ_{\mu}]\big) + R_g(\pa_{\pm}\fJ_{\mu})\,,
    \end{align}
the off-shell flatness condition of $j_{\pm}$\,, and the mCYBE \eqref{mod-YB}. After all, the flatness condition of $\tilde{j}_{\pm}$ \eqref{flatness-tilde-eta} is satisfied under the commutation relations \eqref{tilde-comm1-eta} and \eqref{tilde-comm2-eta}, and the equation of motion \eqref{J-conservation-eta}.

\subsection{The master formula}

By substituting the Lax form \eqref{eta-Lax}, the master formula \eqref{master-formula} is obtained as  
\begin{align}
    S_\eta[g] =-\int_{\cM}d\sigma^+\wedge d\sigma^-\tr (j^\mu\fJ_\mu) =\frac{1}{2}\int_{\cM}d\sigma^+\wedge d\sigma^- \tr(j_-\fJ_+ + j_+\fJ_-)\,,
    \label{eta-action}
\end{align}
which takes exactly the same form as \eqref{YB-action}, though some quantities depend on $c$ and the $R$-operator satisfies the mCYBE \eqref{mod-YB}. At this stage, the homogeneous YB deformation case is reproduced by simply setting $c=0$\,. Hence, we will use the resulting expressions obtained above in the following.


\section{The true Lagrangian}

In the first place, let us summarize the results so far. 

\medskip 

For the YB deformations for any $c$\,, 
the master formula is commonly given by  
\begin{align}
    S_\eta[g]
    &=
    - \int_{\cM}d\sigma^+\wedge d\sigma^-\tr (j^\mu\fJ_\mu)\,,
    \label{identical}
\end{align}
and the three conditions 
\begin{align}
\pa_{+}\fJ_- + \pa_{-}\fJ_+\,=&\, 0\,,
    \label{cond1} \\  
    [\tilde{j}_{+},\fJ_{-}] =&\, [\fJ_{+},\tilde{j}_-]\,,
    \label{cond2}
    \\
    [\tilde{j}_+,\tilde{j}_-]=&\,[\fJ_+,\fJ_-]
    \label{cond3} 
\end{align}
should be satisfied. The deformed current is given by  
\begin{align}
    \label{j-tilde-and-J}
    \tilde{j}_{\pm} =&\, j_{\pm} \pm \eta R_g \,\fJ_{\pm}\,
\end{align}
and the associated Lax form is 
\begin{align}
    \fL_{\pm}=&\, \big(1-c^2\eta^2\big)\,
    \frac{\tilde{j}_{\pm}\pm z\,\fJ_{\pm}}{1-z^2}\,.
    \label{Lax}
\end{align}
In the following, we will refer to the equation numbers from the summary above.  

\medskip 

So far, $\mathfrak{J}_{\pm}$ have not yet been determined, though the constraints to be satisfied are given. 

\subsection{Current expansion}

For the latter analysis, let us expand $\mathfrak{J}_{\mu}$ as
\begin{align}
    \mathfrak{J}_{\mu} 
    =&\,
    2 f_1(\tilde{x}_1,\tilde{x}_2)\,\tilde{j}_\mu + 4f_2(\tilde{x}_1,\tilde{x}_2)\, \tilde{j}_\nu \tr ( \tilde{j}^\nu \tilde{j}_\mu)\,,
    \label{tilde-current}
    \\
    \tilde{x}_1:=&\,
    \tr (\tilde{j}^\mu\tilde{j}_\mu)\,,
    \qquad
    \tilde{x}_2:=
    \tr (\tilde{j}^\mu\tilde{j}_\nu)
    \tr (\tilde{j}^\nu\tilde{j}_\mu)\,,
    \label{tilde-x}
\end{align}
where $f_1$ and $f_2$ are functions subject to satisfying the conditions \eqref{cond3}. The numerical coefficients 2 and 4 are taken for later convenience. Note that the Lorentz scalars should be constructed from $\tilde j_\mu$ rather than $j_\mu$ in order for \eqref{cond2} to be automatically satisfied.

\medskip

For simplicity, suppose that $f_1$ and $f_2$ are specified by a single function $\cF$ as
\al
{
    \label{assumption-f1-and-f2}
    f_1=\partial_1\cF,\qquad
    f_2=\partial_2\cF\,.
}
Then, the current \eqref{tilde-current} depends only on $\mathcal{F}$ like  
\begin{align}
    \fJ_{\mu}=&\,
    2\pa_1 \cF(\tilde x_1, \tilde x_2)\, \tilde{j}_\mu
    +4 \pa_{2}\cF(\tilde x_1, \tilde x_2)\,\tilde{j}_\nu\tr(\tilde{j}^\nu\tilde{j}_\mu)\,, 
    \label{J-expansion}
\end{align}
and 
the constraint \eqref{cond3} is rewritten as
\begin{align}
    4(\partial_{1}\mathcal{F}+\tilde{x}_1\partial_2\mathcal{F})^2 -
	4(2\tilde{x}_2-\tilde{x}_1^2)(\partial_2\mathcal{F})^2 = 1\,.
    \label{PDE-x}
\end{align}
Note that under the current expansion (\ref{J-expansion})\,, there remain two constraints (\ref{cond1}) and (\ref{PDE-x})\,.

\subsection{The correction term}

In the following, let us consider the on-shell condition (\ref{cond1})\,. To this end, it is helpful to use Cartesian coordinates rather than light-cone coordinates. We begin with the following Lagrangian $\mathcal{L}_{\eta}$  
\al
{
	\label{YB-CH-action}
	S_\eta[g]
    = \int\! d\tau\wedge d\sigma\,
    \mathcal{L}_\eta,
    \quad
	\mathcal{L}_\eta
    :=
	\frac{1}{2}\,\tr (\fJ^\mu j_\mu)\,.
}
Taking a variation of $\mathcal{L}_{\eta}$
leads to 
\begin{align}
	\delta \cL_\eta
	=\,
	\frac{1}{2}\tr (\delta \fJ^\mu j_\mu) + \frac{1}{2}\tr (\fJ^\mu \delta j_\mu)\,.
	\label{eta-variation}
\end{align}
The first term on the right-hand side is evaluated as
\al
{
	\frac{1}{2}\tr (\delta \fJ^\mu j_\mu)
	&
	= \frac{1}{2}\tr (\delta \fJ^\mu \tilde j_\mu)
	+ \frac{1}{2} \epsilon_{\mu\nu} \tr (\delta \fJ^\mu \eta R_g\fJ^\nu)
	\no\\[5pt]
	&
	\label{variation-1stTerm}
	= \frac{1}{2}\tr (\delta \fJ^\mu \tilde j_\mu)
	+ \delta
	\biggl(
	\frac{1}{4} \epsilon_{\mu\nu} \tr (\fJ^\mu \eta R_g\fJ^\nu)
	\biggr)
	- \frac{1}{4} \epsilon_{\mu\nu} \tr (\fJ^\mu \eta\, (\delta R_g)\, \fJ^\nu)\,,
}
where the first equality follows from the covariant form of $\tilde{j}$ in \eqref{j-tilde-and-J} 
\als
{
	\label{covariant-j-tilde}
	\tilde j_\mu = j_\mu - \epsilon_{\mu\nu} \eta R_g\fJ^\nu\,,
}
and the second equality uses the fact that $R_g$ is skew-symmetric. Here $\epsilon_{\mu\nu}$ is the anti-symmetric tensor normalized as  $\epsilon^{\tau\sigma}=1$\,.

\medskip

Then, by using $\mathfrak{J}_{\mu}$ in \eqref{J-expansion}, the first term on the right-hand side in \eqref{variation-1stTerm} can be rewritten as
\als
{
	\frac{1}{2}
    \tr (\delta \fJ^\mu \tilde j_\mu)
	& =
	\delta \biggl(\tilde x_1 \partial_1\cF(\tilde x_1,\tilde x_2)
	+ 2 \tilde x_2 \partial_2\cF(\tilde x_1,\tilde x_2) - \frac{1}{2}\cF(\tilde x_1,\tilde x_2)\biggr)\,.
}
This resulting form comes from the simplification in \eqref{assumption-f1-and-f2}. For the detailed computation, see Section 3 of \cite{Fukushima:2025tlj}.

\medskip

The third term on the right-hand side in \eqref{variation-1stTerm} is evaluated as
\al
{
	- \frac{1}{4} \epsilon_{\mu\nu}
    \tr (\fJ^\mu \eta \,(\delta R_g)\, \fJ^\nu)
	& =
	- \frac{1}{4} \epsilon_{\mu\nu}
    \tr (\fJ^\mu [\eta R_g, \text{ad}_{g^{-1}\delta g}]\fJ^\nu)
	\no\\[5pt]
	& =
	- \frac{1}{2} \epsilon_{\mu\nu} \tr ([\fJ^\mu, \eta R_g(\fJ^\nu)] g^{-1}\delta g)
	\no\\[5pt]
	& =
	- \frac{1}{2} \tr ([\fJ^\mu, j_\mu] g^{-1}\delta g)
	\no\\[5pt]
	& =
	- \frac{1}{2} \tr (\fJ^\mu \delta j_\mu)
	- \frac{1}{2} \tr (\partial^\mu \fJ_\mu) g^{-1}\delta g
	+ \frac{1}{2} \partial^\mu (\tr (\fJ_\mu g^{-1}\delta g))\,.
}
Here, we have used the identity 
\begin{eqnarray}
\delta R_g =[R_g, \text{ad}_{g^{-1}\delta g}] 
\qquad \mbox{where} ~~ \text{ad}_{\mathsf{x}}\mathsf{y}:=[\mathsf{x},\mathsf{y}] \quad (^\forall\mathsf{x},\mathsf{y}\in\mathfrak{g})
\end{eqnarray}
in the first equality. 
The second equality follows from the fact that $R_g$ is skew-symmetric, and the third one follows from \eqref{covariant-j-tilde}, together with the covariant form of the off-shell condition \eqref{cond2}. Finally, the fourth equality holds due to $\delta j_\mu = - g^{-1} \delta g \, j_\mu + g^{-1} \partial_\mu \delta g$\,.

\medskip

In summary, the variation of $\mathcal{L}_\eta$ has been evaluated as 
\als
{
	\delta \cL_\eta
	&=
    - \frac{1}{2}
    \tr(\partial^\mu \fJ_\mu g^{-1}\delta g)
    \\
	&\quad
    + \delta \biggl(\tilde x_1 \partial_1\cF(\tilde x_1,\tilde x_2)
	+ 2 \tilde x_2 \partial_2\cF(\tilde x_1,\tilde x_2) - \frac{1}{2}\cF(\tilde x_1,\tilde x_2)
	+ \frac{1}{4} \epsilon_{\mu\nu}
    \tr(\fJ^\mu \eta R_g\fJ^\nu)\biggr)\,, 
}
up to the total derivative terms. Note here that the first term gives the desired equation of motion \eqref{cond1},
while the remaining terms indicate that $\mathcal{L}_{\eta}$ needs to be modified.
Thus, the true Lagrangian $\hat{\mathcal{L}}_\eta$ should be given by 
\als
{
	\label{modified-Lagrangian}
	\frac{1}{2}\hat \cL_\eta
	:=
	\cL_\eta - \tilde x_1 \partial_1\cF(\tilde x_1,\tilde x_2) - 2 \tilde x_2 \partial_2\cF(\tilde x_1,\tilde x_2) + \frac{1}{2}\cF(\tilde x_1,\tilde x_2)
	- \frac{1}{4} \epsilon_{\mu\nu}
    \tr (\fJ^\mu \eta R_g\fJ^\nu)\,, 
}
which is consistent with the Lax form \eqref{Lax}.

\medskip

The true Lagrangian $\hat{\mathcal{L}}_\eta$ has two interpretations.
For the former, by rewriting the original Lagrangian $\mathcal{L}_\eta$ as
\als
{
    \cL_\eta
	&
	= \frac{1}{2}\tr (\fJ^\mu j_\mu)
	= \frac{1}{2}\tr (\fJ^\mu \tilde j_\mu)
	+ \frac{1}{2} \epsilon_{\mu\nu}
    \tr (\fJ^\mu \eta R_g\fJ^\nu)
	\no\\[5pt]
	&= \tilde{x}_1\partial_1\cF(\tilde x_1,\tilde x_2) + 2\tilde{x}_2 \partial_2\cF(\tilde x_1,\tilde x_2)
	+ \frac{1}{2} \epsilon_{\mu\nu}
    \tr (\fJ^\mu \eta R_g\fJ^\nu)
}
and substituting this into \eqref{modified-Lagrangian}, we find that 
\als
{
	\label{corrected-Lagrangian}
    \hat \cL_\eta
	= \cF(\tilde x_1,\tilde x_2)
    + \frac{1}{2} \epsilon_{\mu\nu}
    \tr (\fJ^\mu \eta R_g\fJ^\nu)\,.
}
When $\eta =0$\,, the true Lagrangian $\hat{\mathcal{L}_{\eta}}$ is represented by the undetermined  function $\mathcal{F}$ itself, as shown in \cite{Fukushima:2025tlj}\,. It should be remarked that an additional term proportional to $\eta$ is contained in (\ref{corrected-Lagrangian}) due to the YB deformation.

\medskip

For the latter, $\hat{\mathcal{L}}_\eta$ can also be rewritten as
\als
{
	\hat \cL_\eta
	&= 2\cL_\eta - 2\tilde x_1 \partial_1\cF(\tilde x_1,\tilde x_2)
	- 4 \tilde x_2 \partial_2\cF(\tilde x_1,\tilde x_2)
	+ \cF(\tilde x_1,\tilde x_2)
	- \frac{1}{2} \epsilon_{\mu\nu}
    \tr (\fJ^\mu \eta R_g\fJ^\nu)
	\\[5pt]
	&=
	\cL_\eta - \tilde x_1 \partial_1\cF(\tilde x_1,\tilde x_2)
	- 2 \tilde x_2 \partial_2\cF(\tilde x_1,\tilde x_2)
	+ \cF(\tilde x_1,\tilde x_2)\,.
}
By using the trace of the energy-momentum tensor for $\hat{\mathcal{L}}_\eta$ given by\footnote{The energy-momentum tensor for $\hat{\mathcal{L}}_\eta$ is computed in Appendix \ref{AA}. }
\als
{
    \label{trace-of-EMtensor}
	\hat T^\mu{}_\mu
	= 2(- \tilde x_1 \partial_1\cF(\tilde x_1,\tilde x_2)
	- 2 \tilde x_2 \partial_2\cF(\tilde x_1,\tilde x_2)
	+ \cF(\tilde x_1,\tilde x_2))\,,
}
the true Lagrangian $\hat{\mathcal{L}}_{\eta}$ is expressed as 
\als
{
	\hat \cL_\eta
	=
	\cL_\eta + \frac{1}{2}\hat T^\mu{}_\mu\,.
    \label{trace-correction}
}
That is, the master formula Lagrangian $\mathcal{L}_{\eta}$ should be corrected by the trace of the energy-momentum tensor. 
This expression is the same as in the case with no YB deformation. This result indicates that the correction term is universal, at least for the YB deformations.

\section{The CH construction}\label{sec:introduction}

In the previous section, we have derived the true Lagrangian $\hat{\mathcal{L}}_{\eta}$ that is consistent with the associated Lax form. We shall here determine the explicit form of $\mathcal{F}$ satisfying the PDE (\ref{PDE-x})\,.  

\subsection{The general solution}

The general solution to the PDE \eqref{PDE-x} was given by Courant and Hilbert in \cite{Courant}. To present this solution, it is convenient to introduce new variables $\tilde u$ and $\tilde v$ defined by\footnote{These are similar to the undeformed case \cite{Babaei-Aghbolagh:2025uoz}, but $x_{1}$ and $x_2$ are replaced by $\tilde{x}_{1}$ and $\tilde{x}_2$\,.}
\als
{
	&
	\tilde u := \frac{1}{4}({\textstyle\sqrt{2\tilde x_2 - \tilde x_1^2} - \tilde x_1})\,, \qquad 
	\tilde v := \frac{1}{4}({\textstyle\sqrt{2 \tilde x_2 - \tilde x_1^2} + \tilde x_1})\,.
}
In terms of these variables, the PDE (\ref{PDE-x}) takes a simple form
\als
{
	\label{deformed-CH}
	\partial_{\tilde u}\mathcal{F}\, \partial_{\tilde v}\mathcal{F} = - 1\,.
}
The solution to this equation is given by
\al
{
	\label{CH-solution}
	\mathcal{F}(\tilde u,\tilde v) := \ell(\tau) - \frac{2\tilde u}{\ell'(\tau)}\,,\qquad
	\tau = \tilde v + \frac{\tilde u}{\ell'(\tau)^2}\,, \qquad \ell' := \frac{d\ell}{d\tau}\,.
}
Here $\ell(\tau)$ is an arbitrary function of $\tau$\,, which should satisfy a boundary condition 
\[
\mathcal{F}(0,\tilde v)=\ell(\tilde v)\,. 
\] 
Thus, one can obtain an explicit solution to \eqref{deformed-CH} by choosing a function $\ell$ and then determining $\tau$ algebraically as a function of $\tilde u$ and $\tilde v$\,. 

\medskip 

However, the story has not yet ended in the case with $\eta \neq 0$\,. Note that the arguments of $\mathcal{F}$ are $\tilde{x}_1$ and $\tilde{x}_2$\,, which are composed of $\tilde{j}_{\mu}$ as in (\ref{tilde-x})\,. The expression of $\tilde{j}_{\mu}$ in \eqref{j-tilde-and-J} contains $\mathfrak{J}_{\mu}$\,.
Since $\mathfrak{J}_{\mu}$ contains $\tilde{j}$ as in \eqref{J-expansion}\,,  $\tilde{j}_{\mu}$ in (\ref{j-tilde-and-J}) contains $\tilde{j}_{\mu}$ again! Hence, the situation is quite intricate, and we still need to solve $\tilde{j}_{\mu}$ in terms of $j_{\mu}$\,. 

\subsubsection*{The original YB-sigma model}

Before going to see how to solve $\tilde{j}_{\mu}$ in detail for the general case, it is worth examining the simplest case with no CH-deformation: the original YB-sigma model.  

\medskip 

This case corresponds to 
\begin{equation}
\ell(\tau)=\tau\,.
\end{equation}
Then, the function $\cF$ is given by
\begin{align}
    \label{undeformed-F}
    \cF=\tau - 2\tilde{u} =  \tilde{v}-\tilde{u} = \frac{1}{2}\tilde{x}_1\,.
\end{align}
Substituting this form of $\mathcal{F}$ into \eqref{J-expansion} leads to 
\als
{
    \fJ_{\pm} = \tilde{j}_{\pm}\,.
}
Then, by combining this with \eqref{j-tilde-and-J}, we find that
\als
{
    \tilde{j}_{\pm}
    = \frac{1}{1\mp \eta R_g }j_{\pm}\,.
}
From \eqref{trace-of-EMtensor} with 
\eqref{undeformed-F}\,, we see that the trace of the energy-momentum tensor vanishes: 
\als
{
    \hat T^\mu{}_\mu = 0\,.
}
We thus obtain the true Lagrangian as\footnote{Here, we have assumed that the operator $1-\eta R$ is invertible.}
\begin{align}
    \hat{\cL}_\eta
    =
    \cL_\eta
    =
    - \frac{1}{2}\tr\bigg(j_-\frac{1}{1-\eta R_g}j_+\bigg)\,.
\end{align}
This is indeed the Lagrangian of the YB-deformed PCM.

\subsection{Inversion from \texorpdfstring{$\tilde{j}_{\mu}$}{jmu-tilde} to \texorpdfstring{$j_{\mu}$}{jmu}}

In the following, let us discuss how to solve $\tilde j_\mu$ in terms of $j_\mu$\,.

\medskip 

By combining (\ref{j-tilde-and-J}) and (\ref{J-expansion}), the current $j_{\mu}$ is expressed as 
\als
{
	\label{tilde-j-expression}
	j_\pm
	&=
	(1 \mp 2(\pa_1\mathcal{F}(\tilde x_1,\tilde x_2)
	+ \tilde x_1 \pa_2\mathcal{F}(\tilde x_1,\tilde x_2))\eta R_g) \tilde j_\pm
	\pm 2\,\mathrm{tr}(\tilde j_\pm \tilde j_\pm)\pa_2\mathcal{F}(\tilde x_1,\tilde x_2)\,
	\eta R_g \tilde j_\mp\,.
}
These equations can be written in matrix form as
\als
{
    \label{matrix-form}
	\left(
	\begin{array}{cc}
	j_+ \\[5pt]
	j_-
	\end{array}
	\right)
	=
	\left(
	\begin{array}{cc}
	1 - 2(\pa_1\mathcal{F}
	+ \tilde{x}_1 \pa_2\mathcal{F})\eta R_g
	&
	2\,\mathrm{tr}(\tilde j_+ \tilde j_+)\pa_2\mathcal{F}\eta R_g
	\\[5pt]
	- 2\,\mathrm{tr}(\tilde j_- \tilde j_-)\pa_2\mathcal{F}\eta R_g
	&
	1 + 2(\pa_1\mathcal{F}
	+ \tilde{x}_1 \pa_2\mathcal{F})\eta R_g
	\end{array}
	\right)
	\left(
	\begin{array}{cc}
	\tilde j_+ \\[5pt]
	\tilde j_-
	\end{array}
	\right)\,. 
}
The $2\times 2$ matrix on the right-hand side contains the operator $R_g$ but all elements of the matrix commute with each other. Hence, its determinant is readily evaluated as 
\als
{
    \label{determinant}
    1 - 4(\pa_1\mathcal{F}
	+ \tilde{x}_1 \pa_2\mathcal{F})^2\eta^2 R_g^2
    + 4\,\mathrm{tr}(\tilde j_+ \tilde j_+)\,\mathrm{tr}(\tilde j_- \tilde j_-)
    (\pa_2\mathcal{F})^2\eta^2 R_g^2
    =
    1 - \eta^2 R_g^2\,,
}
where we have used the PDE \eqref{PDE-x}\,, together with
\als
{
	\label{++--product}
    \mathrm{tr}(\tilde j_+ \tilde j_+)\, \mathrm{tr}(\tilde j_- \tilde j_-)
	=
	2\tilde x_2 - \tilde x_1^2\,.
}
From the assumption that $1-\eta R$ is invertible, we can show that $1-\eta^2 R_g^2 $ is invertible. Therefore, the matrix in (\ref{matrix-form}) is invertible, and
the relation \eqref{matrix-form} can be inverted as
\als
{
	\left(
	\begin{array}{cc}
	\tilde j_+ \\[5pt]
	\tilde j_-
	\end{array}
	\right)
	=
	\frac{1}{1 - \eta^2 R_g^2}
	\left(
	\begin{array}{cc}
	1 + 2(\pa_1\mathcal{F}
	+ \tilde{x}_1 \pa_2\mathcal{F})\eta R_g
	&
	- 2\,\mathrm{tr}(\tilde j_+ \tilde j_+)\pa_2\mathcal{F}\eta R_g
	\\[5pt]
	2\,\mathrm{tr}(\tilde j_- \tilde j_-)\pa_2\mathcal{F}\eta R_g
	&
	1 - 2(\pa_1\mathcal{F}
	+ \tilde{x}_1 \pa_2\mathcal{F})\eta R_g
	\end{array}
	\right)
	\left(
	\begin{array}{cc}
	j_+ \\[5pt]
	j_-
	\end{array}
	\right)\,.
}
For later convenience, it is useful to express the above relations as 
\als
{
	\label{tilde-MC-form}
	\tilde j_\pm
	&=
	(O^{(S)} \pm 2(\pa_1\mathcal{F}(\tilde x_1,\tilde x_2) + \tilde x_1 \pa_2\mathcal{F}(\tilde x_1,\tilde x_2))O^{(A)}) j_\pm
    \\[5pt]
	&\quad
    \mp 2\,\mathrm{tr}(\tilde j_\pm \tilde j_\pm)\pa_2\mathcal{F}(\tilde x_1,\tilde x_2) O^{(A)} j_\mp\,,
}
where we have defined new quantities: 
\als
{
	O^{(S)} := \frac{1}{1 - \eta^2 R_g^2}\,,\qquad
	O^{(A)} := \frac{\eta R_g}{1 - \eta^2 R_g^2}\,.
}
The equations \eqref{tilde-MC-form} can be regarded as self-consistency equations for $\tilde j_\mu$\,. Hence, solving them allows us to determine $\tilde j_\mu$ explicitly. Then, putting the resulting $\tilde j_\mu$ into \eqref{J-expansion}\,, we can determine $\mathfrak{J}_\mu$\,. 

\medskip

We can further reduce the self-consistency equations \eqref{tilde-MC-form} to a set of equations for two scalar quantities. First, as shown in Appendix \ref{derivation-of-++--scalars}, one finds that \eqref{tilde-MC-form} leads to
\als
{
	\label{++--scalars}
	\mathrm{tr}(\tilde j_\pm \tilde j_\pm)
	=
	\sqrt{\frac{\mathrm{tr}(O_\pm j_\pm O_\pm j_\pm)}{\mathrm{tr}(O_\mp j_\mp O_\mp j_\mp)}}
	{\textstyle\sqrt{2 \tilde x_2 - \tilde x_1^2}}\,,
}
where the operators $O_\pm$ are defined by
\al{
    \label{pm-operator}
    O_\pm:=\frac{1}{1 \pm \eta R_g}\,.
}
It is convenient to introduce two scalar functions
\al
{
	&
	\label{def-alpha}
	\alpha := \pa_1\mathcal{F}(\tilde x_1,\tilde x_2) + \tilde x_1 \pa_2\mathcal{F}(\tilde x_1,\tilde x_2)\,,
	\\[5pt]
	&
	\label{def-beta}
	\beta := {\textstyle\sqrt{2 \tilde x_2 - \tilde x_1^2}}\,\pa_2\mathcal{F}(\tilde x_1,\tilde x_2)\,.
}
Note that $\alpha$ and $\beta$ satisfy
\als
{
	\label{alpha-beta-relation}
	4\alpha^2 - 4\beta^2 = 1
}
from the PDE \eqref{PDE-x}. In terms of $\alpha$ and $\beta$\,, $\tilde j_\mu$ in \eqref{tilde-MC-form} can be rewritten as
\als
{
	\label{tilde-MC-form-3}
	\tilde j_\pm
	= (O^{(S)} \pm 2\alpha O^{(A)}) j_\pm
    \mp 2 \sqrt{\frac{\mathrm{tr}(O_\pm j_\pm O_\pm j_\pm)}{\mathrm{tr}(O_\mp j_\mp O_\mp j_\mp)}} \beta O^{(A)} j_\mp\,,
}
with the help of \eqref{++--scalars}.

\medskip

By using \eqref{tilde-MC-form-3}, we obtain
\al
{
	\label{++--expressions}
	\mathrm{tr}(\tilde j_\pm \tilde j_\pm) = f_\pm(\alpha,\beta)\,,
    \qquad
	\mathrm{tr}(\tilde j_+ \tilde j_-) = g(\alpha,\beta)\,,
}
where we have defined the following quantities: 
\al
{
	\label{f+-expression}
	f_\pm(\alpha,\beta)
	&:=
	\mathrm{tr}(j_\pm (O^{(S)})^2 j_\pm)
	- 4\, \mathrm{tr}(j_\pm (O^{(A)})^2  j_\pm) \alpha^2
    - 4 \frac{\mathrm{tr}(O_\pm j_\pm O_\pm j_\pm)}{\mathrm{tr}(O_\mp j_\mp O_\mp j_\mp)}\,
	\mathrm{tr}(j_\mp (O^{(A)})^2 j_\mp) \beta^2
	\nonumber\\[5pt]
	& \hspace{-0.2cm}
	- 4 \sqrt{\frac{\mathrm{tr}(O_\pm j_\pm O_\pm j_\pm)}{\mathrm{tr}(O_\mp j_\mp O_\mp j_\mp)}}\,
	\mathrm{tr}(j_+ O^{(S)}O^{(A)} j_-) \beta
	+ 8 \sqrt{\frac{\mathrm{tr}(O_\pm j_\pm O_\pm j_\pm)}{\mathrm{tr}(O_\mp j_\mp O_\mp j_\mp)}}\,
	\mathrm{tr}(j_+ (O^{(A)})^2  j_-) \alpha\beta\,,
	\\[5pt]
	\label{g-expression}
	g(\alpha,\beta)
	&:=
	\mathrm{tr}(j_+ (O^{(S)})^2 j_-)
	- 4\, \mathrm{tr}(j_+ O^{(S)}O^{(A)} j_-) \alpha
	+ 4\, \mathrm{tr}(j_+ (O^{(A)})^2  j_-) (\alpha^2 + \beta^2)
	\nonumber\\[5pt]
	& \hspace{-0.2cm}
	- 4 \sqrt{\frac{\mathrm{tr}(O_-j_- O_-j_-)}{\mathrm{tr}(O_+j_+ O_+j_+)}}\,
	\mathrm{tr}(j_+ (O^{(A)})^2  j_+) \alpha\beta
	- 4 \sqrt{\frac{\mathrm{tr}(O_+j_+ O_+j_+)}{\mathrm{tr}(O_-j_- O_-j_-)}}\,
	\mathrm{tr}(j_- (O^{(A)})^2  j_-) \alpha\beta\,.
}
With these expressions, $\tilde x_1$ and $\tilde x_2$ defined in \eqref{tilde-x} can be written in terms of $\alpha$ and $\beta$ as
\al
{
	\tilde x_1
	= g(\alpha,\beta)\,,\qquad
	\tilde x_2
	= \frac{1}{2} (g(\alpha,\beta))^2 + \frac{1}{2} f_+(\alpha,\beta) f_-(\alpha,\beta)\,.
}
By putting these expressions into \eqref{def-alpha} and \eqref{def-beta}, we arrive at
\al
{
	&
	\label{self-eq-alpha}
	\alpha = \partial_1\mathcal{F}(\tilde x_1(\alpha,\beta),\tilde x_2(\alpha,\beta))
	+ g(\alpha,\beta)\, \partial_2\mathcal{F}(\tilde x_1(\alpha,\beta),\tilde x_2(\alpha,\beta))\,,
	\\[5pt]
	&
	\label{self-eq-beta}
	\beta = {\textstyle\sqrt{f_+(\alpha,\beta)f_-(\alpha,\beta)}}\,
	\partial_2\mathcal{F}(\tilde x_1(\alpha,\beta),\tilde x_2(\alpha,\beta))\,. 
}
These are self-consistency equations for $\alpha$ and $\beta$\,. By solving the algebraic equations \eqref{self-eq-alpha} and \eqref{self-eq-beta} for a given CH solution $\mathcal{F}$, one can determine $\alpha$ and $\beta$ in terms of $j_\mu$\,. Finally, by putting the resulting $\alpha$ and $\beta$\, into \eqref{tilde-MC-form-3}, $\tilde j_\mu$ can be expressed in terms of $j_\mu$.

\medskip 

Finally, we should note that it is quite difficult in general to solve \eqref{self-eq-alpha} and \eqref{self-eq-beta} algebraically solvable. We will present two solvable examples in the next subsection.


\subsection{Examples}

In the following, we shall present two concrete examples: 1) the root $T\bar{T}$-deformed YB sigma model and 2) the $T\bar{T}$-deformed YB sigma model. 

\subsubsection*{Example 1: the root $T\bar T$-deformed YB-sigma model}

The first example is the root $T\bar{T}$-deformed YB-sigma model. 

\medskip 

The associated function $\ell(\tau)$ is given by 
\als
{
	\ell(\tau) = e^\gamma \tau\,,
}
where $\gamma$ is a dimensionless real parameter that measures the root $T\bar T$-deformation. In this case, the solution \eqref{CH-solution} takes the form
\als
{
	\label{CH-rootTTbar}
	\mathcal{F}(\tilde x_1,\tilde x_2)
	=
	\frac{1}{2}\cosh\gamma\cdot \tilde x_1
	+ \frac{1}{2} \sinh\gamma\cdot\textstyle{\sqrt{2 \tilde x_2 - \tilde x_1^2}}\,.
}
For this solution, we have
\al
{
	\label{CH-rootTTbar-del}
	&
	\pa_1\mathcal{F}(\tilde x_1,\tilde x_2)
	=
	\frac{1}{2}\cosh\gamma
	- \frac{1}{2}\sinh\gamma\cdot\frac{\tilde x_1}{\textstyle{\sqrt{2 \tilde x_2 - \tilde x_1^2}}}\,,
	\\[5pt]
	&
	\pa_2\mathcal{F}(\tilde x_1,\tilde x_2)
	=
	\frac{1}{2}\sinh\gamma\cdot\frac{1}{\textstyle{\sqrt{2 \tilde x_2 - \tilde x_1^2}}}\,.
}
In this case, $\alpha$ and $\beta$ defined in \eqref{def-alpha} and \eqref{def-beta}, respectively, are simply constants as
\al
{
	&
	\label{alpha-beta-rootTTbar}
	\alpha = \frac{1}{2}\cosh\gamma\,, \qquad  
	\beta = \frac{1}{2}\sinh\gamma\,.
}
Thus, we do not have to solve equations anymore. Substituting \eqref{alpha-beta-rootTTbar} into \eqref{tilde-MC-form-3}, we obtain
\al
{
	\label{sol-rootTTbar}
    \tilde j_\pm
	= (O^{(S)} \pm \cosh\gamma\cdot O^{(A)}) \, j_\pm
    \mp \sqrt{\frac{\mathrm{tr}(O_\pm j_\pm O_\pm j_\pm)}{\mathrm{tr}(O_\mp j_\mp O_\mp j_\mp)}} \sinh\gamma\cdot O^{(A)}\, j_\mp\,.
}
By using this expression of $\tilde{j}_{\pm}$\,, the root $T\bar{T}$-deformed YB-sigma model and the associated Lax pair have been obtained.

\subsubsection*{Example 2: the $T\bar T$-deformed YB-sigma model}

The next example is the $T\bar{T}$-deformed YB sigma model. 

\medskip

The function $\ell(\tau)$ is given by 
\als
{
	\ell(\tau) = - \frac{1}{\lambda}(1 - \sqrt{1 + 2\lambda\, \tau}\,)\,,
}
where $\lambda$ is a dimensionful parameter that measures the $T\bar T$-deformation. For this choice, the solution \eqref{CH-solution} becomes
\als
{
	\label{CH-TTbar}
	\mathcal{F}(\tilde x_1,\tilde x_2)
	=
	- \frac{1}{\lambda}
	+ \frac{1}{\lambda} \sqrt{1 + \lambda \,\tilde x_1 + \dfrac{\lambda^2}{2} (\tilde x_1^2 - \tilde x_2)}\,.
}
This CH solution leads to the following quantities: 
\al
{
	&
	\pa_1\mathcal{F}(\tilde x_1,\tilde x_2)
	=
	\frac{1 + \lambda \tilde x_1}{2\sqrt{1 + \lambda \tilde x_1 + \dfrac{\lambda^2}{2} (\tilde x_1^2 - \tilde x_2)}}\,,
	\\
	&
	\pa_2\mathcal{F}(\tilde x_1,\tilde x_2)
	=
	\frac{-\dfrac{\lambda}{2}}{2\sqrt{1 + \lambda \tilde x_1 + \dfrac{\lambda^2}{2} (\tilde x_1^2 - \tilde x_2)}}\,.
}
Then, $\alpha$ and $\beta$ in \eqref{self-eq-alpha} and \eqref{self-eq-beta}, respectively, are expressed as 
\al
{
	\alpha
	& = \frac{1 + \dfrac{\lambda}{2} g(\alpha,\beta)}
	{2\sqrt{1 + \lambda g(\alpha,\beta)
	+ \dfrac{\lambda^2}{4} ((g(\alpha,\beta))^2 - f_+(\alpha,\beta)f_-(\alpha,\beta))}}\,, 
	\\[10pt]
	\beta
	& = \frac{-\dfrac{\lambda}{2}{\textstyle\sqrt{f_+(\alpha,\beta)f_-(\alpha,\beta)}}}
	{2\sqrt{1 + \lambda g(\alpha,\beta)
	+ \dfrac{\lambda^2}{4} ((g(\alpha,\beta))^2 - f_+(\alpha,\beta)f_-(\alpha,\beta))}}\,.
}
By combining these such that the square root factor is canceled out, we obtain
\als
{
	\label{alpha-beta-equation-in-TTbar}
    \beta
	+ \frac{\lambda}{2} \beta g(\alpha,\beta)
	+ \frac{\lambda}{2}\alpha \,\textstyle\sqrt{f_+(\alpha,\beta)f_-(\alpha,\beta)}
	= 0\,.
}
By using 
\begin{equation}
    f_-(\alpha,\beta) = \frac{\mathrm{tr}(O_-j_- O_-j_-)}{\mathrm{tr}(O_+j_+ O_+j_+)} \, f_+(\alpha,\beta)\,, 
\end{equation}
which follows from \eqref{++--scalars} and \eqref{++--expressions}, we can rewrite this equation as
\als
{
	\label{alpha-beta-equation-in-TTbar-2}
    \beta
	+ \frac{\lambda}{2} \beta g(\alpha,\beta)
	+ \frac{\lambda}{2}\sqrt{\frac{\mathrm{tr}(O_-j_- O_-j_-)}{\mathrm{tr}(O_+j_+ O_+j_+)}}\alpha
	f_+(\alpha,\beta)
	= 0\,.
}
Furthermore, using $f_+(\alpha,\beta)$ and $g(\alpha,\beta)$ in \eqref{f+-expression} and \eqref{g-expression}, respectively, together with the condition \eqref{alpha-beta-relation} and the identity \eqref{operator-identity}, we find that
\al
{
	&
	\beta g(\alpha,\beta)
	+ \sqrt{\frac{\mathrm{tr}(O_-j_- O_-j_-)}{\mathrm{tr}(O_+j_+ O_+j_+)}}\alpha
	f_+(\alpha,\beta)
	\no\\[5pt]
	&\quad
	= - \mathrm{tr}(j_+ (O^{(S)})^2 j_-)\,\beta
	+ \sqrt{\frac{\mathrm{tr}(O_-j_- O_-j_-)}{\mathrm{tr}(O_+j_+ O_+j_+)}}
	\mathrm{tr}(j_+ (O^{(S)})^2 j_+)\,\alpha
	\no\\[5pt]
	&\qquad
	+ \, \mathrm{tr}(j_+ (O^{(A)})^2  j_-) 4(\alpha^2 - \beta^2)\beta
	- \sqrt{\frac{\mathrm{tr}(O_-j_- O_-j_-)}{\mathrm{tr}(O_+j_+ O_+j_+)}}
	\mathrm{tr}(j_+ (O^{(A)})^2  j_+) 4(\alpha^2 - \beta^2)\alpha
	\no\\[5pt]
	&\quad
	= - \mathrm{tr}(j_+ ((O^{(S)})^2 - (O^{(A)})^2) j_-)\,\beta
	+ \sqrt{\frac{\mathrm{tr}(O_-j_- O_-j_-)}{\mathrm{tr}(O_+j_+ O_+j_+)}}
	\mathrm{tr}(j_+ ((O^{(S)})^2 - (O^{(A)})^2) j_+)\,\alpha
	\no\\[5pt]
	&\quad
	= - \mathrm{tr}(j_+ O^{(S)} j_-)\,\beta
	+ \sqrt{\frac{\mathrm{tr}(O_-j_- O_-j_-)}{\mathrm{tr}(O_+j_+ O_+j_+)}}
	\mathrm{tr}(j_+ O^{(S)} j_+)\,\alpha
	\no\\[10pt]
	&\quad
	= - \mathrm{tr}(j_+ O^{(S)} j_-)\,\beta
	+ \sqrt{\mathrm{tr}(O_+j_+ O_+j_+)\,\mathrm{tr}(O_-j_- O_-j_-)}\,\alpha\,. 
}
Thus, \eqref{alpha-beta-equation-in-TTbar} is reduced to a linear equation in $\alpha$ and $\beta$~:
\als
{
	\beta
	- \frac{\lambda}{2} \,\mathrm{tr}(j_+ O^{(S)} j_-)\, \beta
	+ \frac{\lambda}{2} \sqrt{\mathrm{tr}(O_+j_+ O_+j_+)\,\mathrm{tr}(O_-j_- O_-j_-)} \,\alpha
	= 0\,.
}
Combining this with \eqref{alpha-beta-relation}, we obtain
\al
{
	\label{alpha-TTbar}
	\alpha
	&=
	\frac{1 - \dfrac{\lambda}{2}\mathrm{tr}(j_+ O^{(S)} j_-)}
	{2\sqrt{\biggl(1 - \dfrac{\lambda}{2}\mathrm{tr}(j_+ O^{(S)} j_-)\biggr)^2
	- \dfrac{\lambda^2}{4}\mathrm{tr}(O_+j_+O_+j_+)\,\mathrm{tr}(O_-j_-O_-j_-)}}\,,
	\\[5pt]
	\label{beta-TTbar}
	\beta
	&=
	\frac{ - \dfrac{\lambda}{2}{\textstyle\sqrt{\mathrm{tr}(O_+j_+O_+j_+)\,\mathrm{tr}(O_-j_-O_-j_-)}}}
	{2\sqrt{\biggl(1 - \dfrac{\lambda}{2}\mathrm{tr}(j_+ O^{(S)} j_-)\biggr)^2
	- \dfrac{\lambda^2}{4}\mathrm{tr}(O_+j_+O_+j_+)\,\mathrm{tr}(O_-j_-O_-j_-)}}\,.
}
Finally, by putting \eqref{alpha-TTbar} and \eqref{beta-TTbar} into \eqref{tilde-MC-form-3}, $\tilde{j}_{\mu}$ has been expressed in terms of $j_{\mu}$~:  
\als
{
	\label{sol-TTbar}
	\tilde j_\pm
	& =
	O^{(S)}\, j_\pm
	\pm \frac{1 - \dfrac{\lambda}{2}\mathrm{tr}(j_+ O^{(S)} j_-)}
	{\sqrt{\biggl(1 - \dfrac{\lambda}{2}\mathrm{tr}(j_+ O^{(S)} j_-)\biggr)^2
	- \dfrac{\lambda^2}{4}\mathrm{tr}(O_+j_+O_+j_+)\mathrm{tr}(O_-j_-O_-j_-)}}
	O^{(A)} j_\pm
	\\
	&\quad
	\mp
	\frac{ - \dfrac{\lambda}{2}\mathrm{tr}(O_\pm j_\pm O_\pm j_\pm)}
	{\sqrt{\biggl(1 - \dfrac{\lambda}{2}\mathrm{tr}(j_+ O^{(S)} j_-)\biggr)^2
	- \dfrac{\lambda^2}{4}\mathrm{tr}(O_+j_+O_+j_+)\mathrm{tr}(O_-j_-O_-j_-)}}
	O^{(A)} j_\mp\,.
}
By using this form of $\tilde{j}_{\mu}$\,, the classical action of the $T\bar{T}$-deformed YB-sigma model and the associated Lax pair have been derived explicitly.

\section{Conclusion and Discussion}

In this paper, we have incorporated the YB deformations into our previous work \cite{Fukushima:2025tlj}. 
As in the case with no YB deformation, the true Lagrangian $\hat{\cL}_\eta$ is represented by the sum of the master formula Lagrangian $\cL_\eta$ and the trace of the energy-momentum tensor $\hat{T}^\mu{}_\mu$\,. This is the same as in the case with no YB deformation. As a result, the correction term is universal at least under the YB deformations.

\medskip

The true Lagrangian can be determined basically by following the CH construction, though the Lagrangian $\hat{\cL}_{\eta}$ has a nested structure. By solving the PDE, the function $\mathcal{F}$ can be determined. However, the arguments of $\mathcal{F}$ are two independent Lorentz scalars $\tilde{x}_1$ and $\tilde{x}_2$ defined in \eqref{tilde-x} and these are composed of $\tilde{j}_{\mu}$\,. Since this $\tilde{j}_{\mu}$ is determined through the self-consistency conditions, we need to solve them and express $\tilde{j}_{\mu}$ in terms of $j_{\mu}$\,. In this paper, we have presented a general scheme to do that. However, it is quite difficult, in general, to carry it out. We have done this for two examples explicitly: the root $T\bar{T}$-deformation and the $T\bar{T}$-deformation. 




\medskip 

There are some future problems. It is significant to further test the universality of the correction term to the master formula Lagrangian. At least so far, the universality\footnote{The correction term in PCM and $\mathcal{E}$-model \cite{Klimcik:2015gba,Klimcik:2017ken} for the $T\bar{T}$-deformation was found in \cite{Sakamoto:2025hwi} and its expression is the same. This result also supports the universality.} has been checked for PCM and its YB-deformations in \cite{Fukushima:2025tlj} and this paper.
Since the symmetric coset sigma models are incorporated into the AFSM framework, it is reasonable to extend the CH and YB deformations to the symmetric coset models by following the work \cite{Fukushima:2020dcp}. Along this line, it is nice to consider bi-YB deformations \cite{Klimcik:2014bta} as well. Such extensions would provide further support for the universality of the correction term. 

\medskip

It is significant to consider the physical conditions on the function $\ell(\tau)$ in systems with CH and YB deformations. The causality and analyticity conditions have been discussed in the context of nonlinear electrodynamics~\cite{Russo:2024llm,Russo:2025fuc,Babaei-Aghbolagh:2025cni,Babaei-Aghbolagh:2026vkm}. It is important to clarify the consequences of the YB deformation for these conditions and the implications for 2D integrable sigma models. 

\medskip

It is also an exciting issue to study the CH and YB deformations in the context of String Theory and the AdS/CFT correspondence \cite{Maldacena:1997re} by considering, for example, AdS$_3$ and S$^3$\,. In particular, it is very fascinating to explore the boundary-theory interpretation of the CH and YB deformation.  
\medskip 

We hope that the unification of the CH and YB deformations could shed new light on the subject of integrable systems.

\subsection*{Acknowledgments}

The work of O.\,F.\ was supported by RIKEN Special Postdoctoral Researchers Program and JSPS Grant-in-Aid
for Research Activity Start-up No.\,24K22890.
The work of K.Y. was supported by MEXT KAKENHI Grant-in-Aid for Transformative Research Areas A “Machine Learning Physics” No.\,22H05115, and JSPS Grant-in-Aid for Scientific Research (B) No.\,22H01217 and (C) No.\,25K07313, and the Asahipen Hikari Foundation.

\appendix

\section*{Appendix}


\section{The energy-momentum tensor for $\hat{\mathcal{L}}_{\eta}$}
\label{AA}

Let us here compute the energy-momentum tensor for the Lagrangian $\hat{\mathcal{L}}_\eta$ given in  \eqref{corrected-Lagrangian}\,.

\medskip

To do this, let us decompose $\hat{\mathcal{L}}_\eta$ like 
\al{
    &
    \hat{\mathcal{L}}_\eta
    = \hat{\mathcal{L}}_\eta^{(G)}
    + \hat{\mathcal{L}}_\eta^{(B)}\,,
    \\[5pt]
    &
    \hat{\mathcal{L}}_\eta^{(G)}
    := \cF(\tilde x_1,\tilde x_2)\,, \qquad
    \hat{\mathcal{L}}_\eta^{(B)}
    := \frac{1}{2} \epsilon_{\mu\nu}
    \tr (\fJ^\mu \eta R_g\fJ^\nu)\,,
    }
where $\hat{\mathcal{L}}_\eta^{(G)}$ and $\hat{\mathcal{L}}_\eta^{(B)}$ are the metric part and the $B$-field part, respectively. Then the energy-momentum tensor for $\hat{\mathcal{L}}_\eta$ is defined as 
\begin{align}
    \label{EM-tensor-Lhat}
    \hat{T}_{\mu\nu}:=&\,
    -2\frac{\pa \hat{\cL}_\eta}{\pa \eta^{\mu\nu}}
    + \eta_{\mu\nu}\hat{\cL}_\eta^{(G)}
    \no\\
    =&\,
    - 2\pa_1\cF(\tilde{x}_1,\tilde{x}_2)\tr (\tilde{j}_\mu\tilde{j}_\nu)
    - 4\pa_2\cF(\tilde{x}_1,\tilde{x}_2)\tr(\tilde{j}_\mu\tilde{j}^\rho)\tr(\tilde{j}_\nu\tilde{j}_\rho)
    +\eta_{\mu\nu}\cF(\tilde{x}_1,\tilde{x}_2)\,. 
\end{align}
Note that $\cL_\eta^{(B)}$ is irrelevant because it is independent of the metric.

\medskip

As a result, the trace of $\hat{T}_{\mu\nu}$ is given by 
\begin{align}
    \hat{T}^{\mu}{}_{\mu}
    =&\,
    - 2\tilde{x}_1\pa_1\cF(\tilde{x}_1,\tilde{x}_2)
    - 4\tilde{x}_2\pa_2\cF(\tilde{x}_1,\tilde{x}_2)
    + 2\cF(\tilde{x}_1,\tilde{x}_2)\,. 
\end{align}

\section{Detailed derivation of \eqref{++--scalars}}
\label{derivation-of-++--scalars}

We shall explain in detail how to derive the equation (\ref{++--scalars})\,, that is, 
\als
{
	\label{B1}
	\mathrm{tr}(\tilde j_\pm \tilde j_\pm)
	=
	\sqrt{\frac{\mathrm{tr}(O_\pm j_\pm O_\pm j_\pm)}{\mathrm{tr}(O_\mp j_\mp O_\mp j_\mp)}}
	{\textstyle\sqrt{2 \tilde x_2 - \tilde x_1^2}}\,.
}

\medskip 

With the help of \eqref{tilde-MC-form}, we can obtain the following expressions: 
\al
{
	\label{tilde++--}
	\mathrm{tr}(\tilde j_\pm \tilde j_\pm)
	&=
	\mathrm{tr}(j_\pm (O^{(S)})^2 j_\pm)
	 - 4(\pa_1\mathcal{F} + \tilde x_1\pa_2\mathcal{F})^2 \,\mathrm{tr}(j_\pm (O^{(A)})^2  j_\pm)
	\nonumber\\[5pt]
	&\quad
	 - 4\,\mathrm{tr}(\tilde j_\pm \tilde j_\pm)\,\mathrm{tr}(\tilde j_\pm \tilde j_\pm)
	 (\pa_2\mathcal{F})^2 \,\mathrm{tr}(j_\mp (O^{(A)})^2 j_\mp)
	\nonumber\\[5pt]
	&\quad
	 - 4\,\mathrm{tr}(\tilde j_\pm \tilde j_\pm)\pa_2\mathcal{F}\,
	 \mathrm{tr}(j_+ O^{(S)}O^{(A)} j_-)
	\nonumber \\[5pt]
	 &\quad
	+ 8\,\mathrm{tr}(\tilde j_\pm \tilde j_\pm)
	\pa_2\mathcal{F} (\pa_1\mathcal{F} + \tilde x_1 \pa_2\mathcal{F})\,
	\mathrm{tr}(j_+ (O^{(A)})^2  j_-)\,,
	\\[10pt]
	\label{tilde-j+j-}
	\mathrm{tr}(\tilde j_+ \tilde j_-)
	&=
	\mathrm{tr}(j_+ (O^{(S)})^2 j_-)
	- 4(\pa_1\mathcal{F} + \tilde x_1\pa_2\mathcal{F})\,\mathrm{tr}(j_+ O^{(S)}O^{(A)} j_-)
	\nonumber\\[5pt]
	&\quad
	+ 4 ((\pa_1\mathcal{F} + \tilde x_1\pa_2\mathcal{F})^2
	+ \,\mathrm{tr}(\tilde j_+ \tilde j_+)\,\mathrm{tr}(\tilde j_- \tilde j_-)(\pa_2\mathcal{F})^2)\,
	\mathrm{tr}(j_+ (O^{(A)})^2  j_-)
	\nonumber\\[5pt]
	&\quad
	- 4\,\mathrm{tr}(\tilde j_- \tilde j_-)\pa_2\mathcal{F} (\pa_1\mathcal{F} + \tilde x_1 \pa_2\mathcal{F})\,
	\mathrm{tr}(j_+ (O^{(A)})^2  j_+)
	\nonumber\\[5pt]
	&\quad
	- 4\,\mathrm{tr}(\tilde j_+ \tilde j_+)\pa_2\mathcal{F} (\pa_1\mathcal{F} + \tilde x_1 \pa_2\mathcal{F})\,
	\mathrm{tr}(j_- (O^{(A)})^2  j_-)\,. 
}
Using \eqref{tilde++--}, we find that
\al
{
	\label{difference-identity}
	0
	&
	= \mathrm{tr}(\tilde j_+ \tilde j_+) \,\mathrm{tr}(\tilde j_- \tilde j_-)
	- \mathrm{tr}(\tilde j_- \tilde j_-)\,\mathrm{tr}(\tilde j_+ \tilde j_+)
	\no\\[5pt]
	&
	= \mathrm{tr}(j_+ (O^{(S)})^2 j_+) \,\mathrm{tr}(\tilde j_- \tilde j_-)
	- \mathrm{tr}(j_- (O^{(S)})^2 j_-)\,\mathrm{tr}(\tilde j_+ \tilde j_+)
	\no\\[5pt]
	&\quad
	- (4(\pa_1\mathcal{F} + \tilde x_1\pa_2\mathcal{F})^2
	- \,\mathrm{tr}(\tilde j_+ \tilde j_+)\,\mathrm{tr}(\tilde j_- \tilde j_-)(\pa_2\mathcal{F})^2)
	\,\mathrm{tr}(j_+ (O^{(A)})^2  j_+)\,\mathrm{tr}(\tilde j_- \tilde j_-)
	\no\\[5pt]
	&\quad
	+ (4(\pa_1\mathcal{F} + \tilde x_1\pa_2\mathcal{F})^2
	- \,\mathrm{tr}(\tilde j_+ \tilde j_+)\,\mathrm{tr}(\tilde j_- \tilde j_-)(\pa_2\mathcal{F})^2)
	\,\mathrm{tr}(j_- (O^{(A)})^2  j_-)\,\mathrm{tr}(\tilde j_+ \tilde j_+)
	\no\\[5pt]
	&
	= \mathrm{tr}(j_+ ((O^{(S)})^2 - (O^{(A)})^2) j_+) \,\mathrm{tr}(\tilde j_- \tilde j_-)
	- \mathrm{tr}(j_- ((O^{(S)})^2 - (O^{(A)})^2) j_-)\,\mathrm{tr}(\tilde j_+ \tilde j_+)
	\no\\[5pt]
	&
	= \mathrm{tr}(j_+ O^{(S)} j_+) \,\mathrm{tr}(\tilde j_- \tilde j_-)
	- \mathrm{tr}(j_- O^{(S)} j_-)\,\mathrm{tr}(\tilde j_+ \tilde j_+)\,.
}
Here, the third equality follows from the PDE \eqref{PDE-x} together with \eqref{++--product},
and the last equality follows from the identity
\als
{
	\label{operator-identity}
	(O^{(S)})^2 - (O^{(A)})^2 = \frac{1}{1 - \eta^2 R_g^2} = O^{(S)}\,.
}
In terms of the operators $O_\pm$ defined in \eqref{pm-operator}, the final expression in \eqref{difference-identity} is rewritten as
\als
{	
	\label{++--relation}
    \mathrm{tr}(O_+j_+ O_+j_+)
	\,\mathrm{tr}(\tilde j_- \tilde j_-)
	=
	\mathrm{tr}(O_-j_- O_-j_-)
	\,\mathrm{tr}(\tilde j_+ \tilde j_+)\,.
}
By combining this relation with \eqref{++--product}, we arrive at \eqref{++--scalars} (equivalently (\ref{B1}))\,.

\bibliographystyle{utphys}
\bibliography{auxiliary}


\end{document}